\documentclass{aa}  

\usepackage{longtable}
\usepackage{graphicx}
\usepackage{txfonts}
\usepackage{longtable,lscape}

\def\R23{\mbox{$\rm R_{23}$}}

\def\kmsmpc{km s$^{-1}$ Mpc$^{-1}$}

\def\Hb{\mbox{${\rm H}{\beta}$}}
\def\Ha{\mbox{${\rm H}{\alpha}$}}
\def\NII{\mbox{${\rm [N\,II]\,}{\lambda\,6584}$}}
\def\NIIbl{\mbox{${\rm [N\,II]\,}{\lambda\,6548}$}}

\begin{document}

\title{Cluster induced quenching of galaxies in the massive cluster XMMXCS\,J2215.9-1738 at $z \sim 1.5$ traced by enhanced metallicities inside half $R_{200}$ 
}


\author{C.~Maier\inst{1}
\and M.~Hayashi\inst{2}
\and B.~L.~Ziegler\inst{1}
\and T.~Kodama\inst{3}
}

\institute{University of Vienna, Department of Astrophysics, Tuerkenschanzstrasse 17, 1180 Vienna, Austria\\
\email{christian.maier@univie.ac.at}
\and National Astronomical Observatory of Japan, Osawa, Mitaka, Tokyo 181-8588, Japan
\and Astronomical Institute, Tohoku University, Aramaki, Aoba-ku, Sendai 980-8578, Japan
}

\titlerunning{Strangulation in a cluster at $z \sim 1.5$}
\authorrunning{C. Maier et al.}

\date{Received ; accepted}

\abstract 
{}
{Cluster environments at $z<0.5$ were found to increase the gas metallicities of galaxies which enter inner regions of the clusters where the density of the intracluster medium is high enough to remove their hot halo gas by ram-pressure stripping effects and to stop the inflow of pristine gas. To extend these studies to $z>1$, the most massive clusters known at these redshifts are the sites where these environmental effects should be more pronounced and more easily observed with present day telescopes.}
{We explore the massive cluster XMMXCS\,J2215.9-1738 at $z \sim 1.5$ with KMOS spectroscopy of \Ha\, and \NII\, covering a region that corresponds to about one virial radius. Using published spectroscopic redshifts of 108 galaxies  in and around the cluster we computed the location of galaxies in the projected velocity-versus-position phase-space to separate our cluster sample into a virialized region of objects accreted longer ago (roughly inside  half $R_{200}$) and a region of infalling galaxies.
We measured oxygen abundances for ten cluster galaxies with detected \NII\, lines in the individual galaxy spectra and compared the mass--metallicity relation of the galaxies inside  half $R_{200}$ with the infalling galaxies and a field sample at similar redshifts.}
{We find that the oxygen abundances of individual $z \sim 1.5$ star-forming cluster galaxies  inside  half $R_{200}$ are comparable, at the respective stellar mass, to the higher local SDSS metallicity values. 
We compare our measurements with a field galaxy sample from the KMOS3D survey at similar redshifts. We find that the \NII/\Ha\, line ratios inside  half $R_{200}$ are higher by 0.2\,dex and that the resultant metallicities of the galaxies in the inner part of the cluster are higher by about 0.1\,dex, at a given mass, than the metallicities of infalling galaxies and of field galaxies at $z \sim 1.5$.
The enhanced metallicities of cluster galaxies at $z \sim 1.5$ inside $0.5 R_{200}$ indicate that the density of the intracluster medium in this massive cluster becomes high enough toward the cluster center such that the ram pressure exceeds the restoring pressure of the hot gas reservoir of cluster galaxies.  This can remove the gas reservoir and initiate quenching; although the galaxies continue to form stars, albeit at slightly lower rates, using the available cold gas in the disk which is not stripped. 
}
{}


\maketitle



\setcounter{section}{0}
\section{Introduction}
\label{sec:intro}
~~~While local and low-redshift clusters host large fractions of passive galaxies, the star-forming (SF) population becomes
larger with increasing redshift. Since clusters of galaxies grow by accreting mass from their surroundings, 
the SF galaxies observed in high-z clusters must be the progenitors of the local passive galaxies.
The $1 < z < 2$ redshift range hosts the emergence of the Hubble
sequence of disks and elliptical galaxies  and the buildup of a
significant fraction of the stellar mass in the universe \citep[e.g.,][]{dickins03,drory05}, which is a transition epoch for clusters. 
At higher redshifts, forming clusters (protoclusters) consisting of mostly SF galaxies are forming a vast majority of their stellar content \citep[e.g.,][]{chiang17}. At later epochs ($z<1$), many massive clusters virialize and most galaxies in their central parts are passive. Therefore, this $1 < z < 2$ transition period when cluster galaxies are at the onset of environmental influences is a crucial time to investigate their properties and the mechanisms driving their evolution.

~~While studies of the mass-metallicity relation (MZR),
chemical evolution, 
and physical conditions of the interstellar medium (ISM) of SF \emph{field} galaxies at these redshifts are now based on larger samples \citep[e.g.,][]{zahid14,kashino17}, the samples of \emph{cluster} SF galaxies at $z>1$ with studied metallicities are still very small.
There is ongoing debate about environmental signatures in the chemical enrichment of cluster galaxies, especially at $z>1$.
The few available samples are small (compared to field samples) and are affected by selection biases and the metallicity calibrator used, 
and produced some contradictory results.
\citet{kulas13} found a 0.15\,dex metallicity enhancement for $z\sim 2.3$ proto-cluster galaxies with respect to field counterparts at lower masses (by stacking galaxies with nondetected [NII] emission lines), 
but no difference at higher masses. 
\citet{shimakawa15} found higher metallicities in 
proto-cluster members at $z>2$ than in the field for $10<\rm{log}(\rm{M/M}_{\odot})<11$, by measuring metallicities based on [NII]/\Ha\, from stacked spectra.
On the other hand, \citet{valent15} found lower [NII]/H$\alpha$ ratios in $z \sim 2$ protocluster galaxies 
than in the field.  
\citet{kacprzak15} and \citet{tran15} used  stacked [NII]/H$\alpha$ ratios (including upper limits for [NII]) and found a similar MZR relation for field and cluster galaxies at $z \sim 2.1$ and $z \sim 1.62$, respectively. 

%

At lower redshifts there is still debate over the amplitude of environmental effects on metallicities, with studies affected by sample selections and metallicity estimators, but also by the definition of environment. 
Defining environment by local densities, \citet{mouhcine07} reported variation in metallicity at a given mass ranging from 0.02\,dex for massive to 0.08\,dex for lower-mass galaxies.
Studying central and satellite SDSS galaxies, \citet{pasq12} and \citet{pengmaio14} both reported an average metallicity of satellites  higher than that of centrals, especially for low-stellar-mass galaxies.
Another investigation of 1318 galaxies in local clusters with a range of halo masses of $\sim 10^{13}-10^{15}M_{\odot}$ found an 
increase of 0.04\,dex in oxygen abundances for galaxies in clusters compared to the field \citep{elli09}.
\citet{cooper08} used the SDSS sample and 
claimed that there is a stronger 
relationship between metallicity and environment (largely driven by galaxies in high-density regions such as groups and clusters), such that more metal-rich galaxies favor regions of higher overdensity.  
On the other hand, \citet{hughes13} found a similar MZR in field galaxies compared to Virgo cluster galaxies with a slight trend of increasing metallicities in cluster galaxies compared to field galaxies at lower masses (their Fig.\,7), but not significant due to their small number statistics.
\citet{gupta16} studied two clusters at $z \sim 0.35$ finding a higher MZR in cluster galaxies compared to field galaxies in one cluster, and no difference in the MZR of field and cluster galaxies in the other cluster. 
As discussed below and driven by our metallicity studies in $z<0.5$ $M_{200} \sim 10^{15}M_{\odot}$ clusters \citep{maier16,maier19}, it seems that environmental effects are stronger in more dense environments at $z<0.5$ and that a higher cluster mass produces stronger effects on metallicities.

%

Most of the studies of the MZR of cluster galaxies at $z>1$ have used the \NII/\Ha\, ratio to determine metallicities with the N2-calibration \citep{petpag04}.
The [NII]/\Ha\, ratio  involves two emission lines (ELs) that are close
to each other in wavelength and has some benefits: it can be observed at the same time with higher-resolution spectroscopy, the ratio is relatively insensitive to extinction correction, and, from the ground, \Ha\, and \NII\, can be followed all the way to redshift $z \sim 2.5$ (the limit of the near-infrared (NIR) K-band).
On the other hand, one main issue with the [NII]/\Ha\, ratio is that the \NII\, line is often too weak to be detected in individual spectra at higher redshifts, therefore the stacking of spectra of several cluster galaxies was often used in the literature to study their chemical enrichment. Another problem is that the contamination of [NII] by night-sky lines can produce spurious metallicities relying only on [NII]/\Ha. This issue was investigated by  \citet{magdis16}, who used K-band Multi Object Spectrograph (KMOS) observations of KROSS (KMOS Redshift One Spectroscopic Survey) galaxies at $z \sim 1$.  
These latter authors found that imperfect subtraction of OH sky lines affected 113 galaxies with no [NII] detection in their sample with distances $<7$\AA\, between the expected central $\lambda$ of \NII\,  and sky lines, producing incorrect [NII] upper limits on flux measurements and spurious metallicities from stacked spectra; they concluded that very weak \NII\, lines partly overlapping
with strong sky lines may not be used to determine reliable metallicities based on using [NII]/\Ha\, only.
In the samples of cluster galaxies at $z>1.6$  with MZR studies mentioned above, it is possible that there are galaxies with no detection of [NII] and a likely problematic measurement of the \NII\, EL flux (from stacking these spectra) due to the OH sky-line residuals at the position of [NII], which could bias the resulting metallicity measurement from stacked spectra.

\begin{figure}
\includegraphics[width=9cm,angle=0,clip=true]{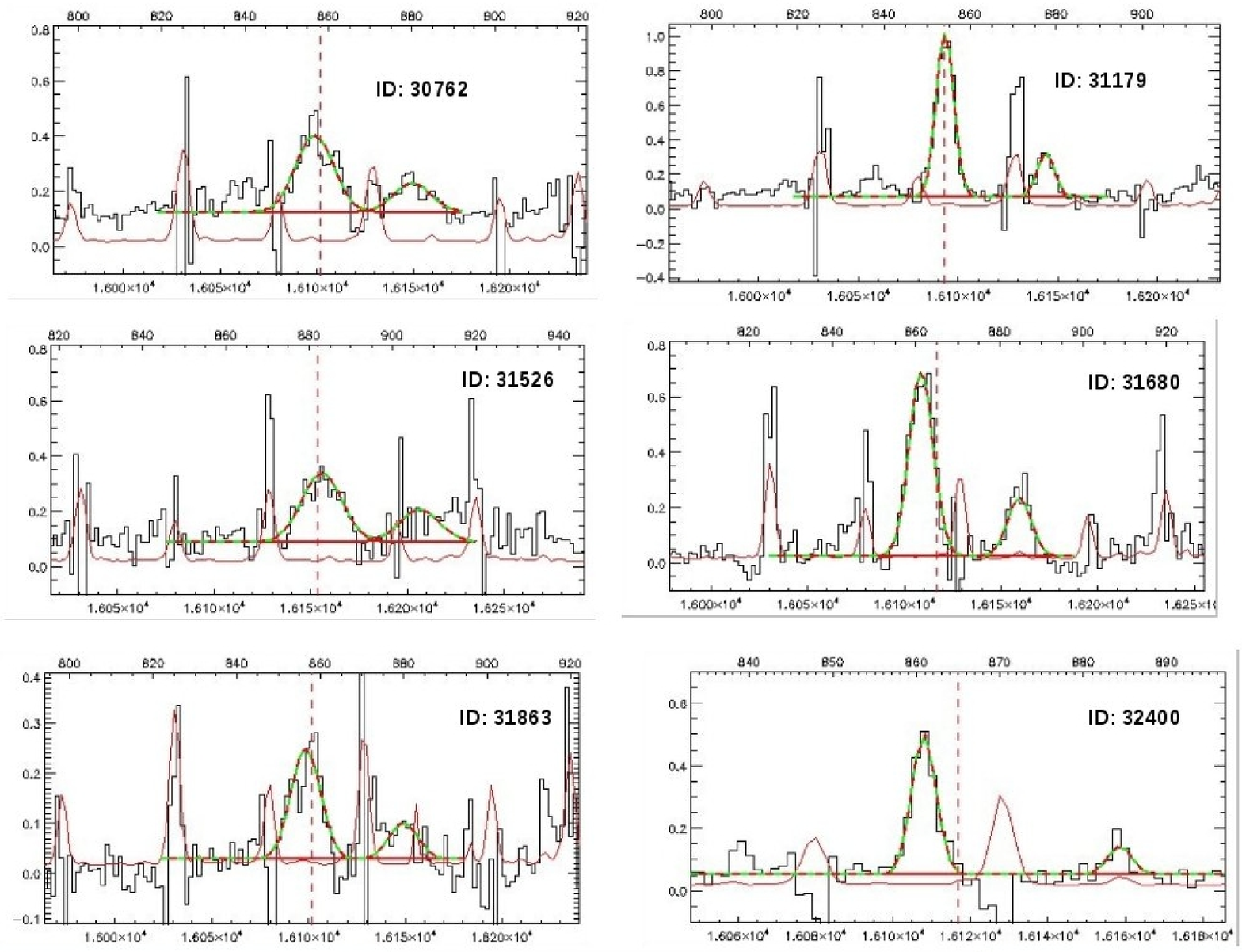}
\caption
{
\label{fig:HaNII} 
\footnotesize 
The six galaxies of the main \emph{SFvirialized} sample. 
The panels show for each galaxy the extracted spectrum (black solid line) over an area of 25 spaxels with the fitted Gaussians by KUBEVIZ to \Ha\, and \NII\, shown in green and the noise spectrum as a solid red line.
It can be seen that the measurements of \Ha\, and \NII\, are not affected by strong sky lines (peaks of the solid red lines).
}
\end{figure}


At lower redshifts, using a sample of SF cluster galaxies at $z \sim 0.2$ and $z \sim 0.4$,
\citet{maier16,maier19} found higher metallicities at a given mass for cluster galaxies in the inner parts of seven massive LoCuSS clusters and one CLASH cluster.
\citet{maier16,maier19} compared their observational findings with metallicity-star-formation rate (SFR)-mass bathtub model predictions with inflowing gas \citep{lilly13} and deduced a slow-quenching (strangulation) scenario in which the gas metallicities can increase after removal of the hot halo gas reservoir because the interstellar medium (ISM) is no  longer diluted by the inflow of pristine gas.
During strangulation, the inflow of new gas is cut out, but the cold ISM disk is not directly perturbed. In this case the star formation can continue, using the gas available in the disk (resulting in higher gas-phase metallicities) until the cold gas is completely used up.

~~~These observational studies  were compared with high-resolution cosmological hydrodynamic simulations of \citet{bahe13}, who computed the density of the intracluster medium (ICM) and derived the ram pressure in clusters.
The conclusion was that the ICM becomes dense enough at $R \sim R_{200}$,
in $z<0.5$ massive clusters with $M_{200}\sim 10^{15}M_{\odot}$,  to remove the {hot} halo gas surrounding
massive galaxies ($10<\rm{log}(\rm{M/M}_{\odot})<11$) by ram-pressure stripping (RPS) effects. 
$R_{200}$ is the radius that encloses a mean density 200 times
the critical density at a given redshift.
For lower-mass galaxies ($9<\rm{log}(\rm{M/M}_{\odot})<10$), the ICM becomes dense enough to remove their
{hot} halo gas  already further out  at $R \sim 2R_{200}$.
\citet{maier19} further deduced that, while slowly quenching for 1-2\,Gyrs, galaxies travel to denser inner regions of the cluster where the RPS
also exceeds the restoring pressure of the cold gas, eventually completely quenching star formation by stripping the {cold} gas in a {rapid} phase.
This {slow-then-rapid} quenching scenario is also in agreement with the theoretical study of \citet{stein16}, who used ram-pressure stripping simulations employing the moving-mesh code AREPO \citep{springel10} to follow at high resolution the interaction of a galaxy cluster with infalling galaxies. \citet{stein16} found that typically their model galaxies continue to form stars with only slightly modified rates as a result of the stripping of the hot gaseous halos (strangulation) of the galaxies. On the other hand, the cold gas of their simulated galaxies is stripped only during pericenter passages with small pericenter distances leading to a full quenching of star formation on a short timescale.


\begin{table*}
\caption{Measured and derived quantities for 19 XMM2215 cluster galaxies at  $z \sim 1.46$ with KMOS H-band observations. 
The complete version of this table (now only three objects) for the full sample of 19 galaxies will be available after the paper is published in A\&A.
}
\label{MeasXMM2215}
\begin{tabular}{ccccccc}
\hline\hline      
Id & spect. z & log[NII]/\Ha &   log(M/M$_{\odot}$) & O/H (PP04) & $R/R_{200}$ \\
\hline  
\vspace{3pt}
29284 &    1.461  &  -0.57 $\pm$  0.08 &  10.36 $^{-0.03}_{+0.03}$  &   8.57 $\pm$ 0.05& 0.74  \\ 
\vspace{3pt}
29609 &    1.459  &      [NII] on OH   &  10.49 $^{-0.08}_{+0.11}$  &      *           & 0.44  \\ 
\vspace{3pt}
30183 &    1.458  &      [NII] on OH   &  10.18 $^{-0.04}_{+0.03}$  &      *           & 0.27  \\ 
\hline                                                                         
\end{tabular}
\end{table*}


~~~As discussed also in \citet{maier19}, there is a significant difference between this {slow-then-rapid} scenario and the ``delayed-then-rapid'' scenario of \citet{wetzel13}. In the ``slow'' quenching phase the galaxies are already affected by the environment because strangulation was inititated, while in the ``delayed'' phase of the \citet{wetzel13} scenario the galaxies are completely unaffected by environment.
A more recent study of \citet{roberts19} used X-ray data for  $z<0.1$ clusters to study
the influence of the ICM on the quenching of satellite galaxies.
They found that the quenched fraction of galaxies increases modestly at ICM densities below a threshold before increasing sharply beyond this
threshold toward the cluster center. 
These latter authors found their results to be consistent with a picture where cluster galaxies experience an initial, slow-quenching mode driven by steady gas depletion, followed by rapid quenching associated with ram pressure of cold-gas stripping near 
the cluster center, in agreement with the \emph{slow-then-rapid} quenching scenario of \citet{maier19}.

These investigations of lower-redshift clusters demonstrated that the density of the ICM is an important parameter influencing the quenching of galaxies and that enhanced metallicities in massive clusters are a valuable quenching tracer. 
To extend these studies to $z>1$ we aim to study clusters of galaxies found as overdensities of galaxies which are unambiguosly identified through their hot diffuse medium, manifested as X-ray emission. 
One of the most massive galaxy clusters at $z>1.5$ is  XMMXCS\,J2215.9-1738 (hereafter XMM2215) at $z \sim 1.46$ with extended X-ray emission from the hot gas, found from XMM Newton X-ray data as part of the XMM Cluster Survey \citep{stanford06}.
Given its high mass and high ICM density possibly favoring stronger RPS, this is one of the best cluster targets to study environmental effects on metallicities and quenching at $z>1$, because more pronounced environmental effects than in other less-massive clusters at $z>1$ are expected.

 A pioneering study of the MZR in XMM2215 was done by \citet{hayashi11}. They claimed to find a similar MZR relation for cluster  and field galaxies at $z \sim 1.5$. However, a caveat of their study was the use of relatively low-resolution ($R \sim 700$) Subaru-MOIRCS spectroscopy which made night-sky line-correction difficult, and therefore  night-sky-line contamination of [NII] influencing metallicity measurements (as mentioned above) could not be always avoided. For most of their observed galaxies the signal-to-noise ratio (S/N) of the [NII] fluxes was relatively low (S/N $<2-3$) or only upper limits for [NII] line fluxes could be measured. 
The new KMOS observations which we present in this study also contain some of the galaxies studied by \citet{hayashi11}. The much better resolution of KMOS ($R \sim 4000$ in H-band vs. $R \sim 700$ with MOIRCS) and better S/N from the higher sensitivity 
Integral Field Unit (IFU) data enables us to improve the S/N of \Ha\, and [NII] in all galaxies in common with \citet{hayashi11}. 
Additionally, we revise the study of the chemical enrichment in XMM2215 by \citet{hayashi11} now additionally using the information from the phase-space diagram about the position of galaxies in the cluster, by separating galaxies from the infalling and virialized regions.

The paper is structured as follows: In Sect. 2 we
present the selection of the XMM2215 cluster EL galaxies at $z \sim 1.5$, their KMOS spectroscopy, and data reduction.
We describe the EL flux measurements 
and present the derivation of SFRs, metallicities, and stellar masses of the observed cluster galaxies.
In Sect.\,3 we present  the phase-space diagram, 
 the mass$-$specific SFR (SSFR) relation, and the MZR at $z\sim 1.5$.
We investigate how the cluster environment affects the chemical enrichment.
In Sect.\,4 
we discuss the slow  quenching scenario in a massive cluster at $z>1$ implied by our findings and the comparison with literature metallicity studies in clusters at $z>1$.
Finally, we summarize our conclusions.
A concordance cosmology with $\rm{H}_{0}=70$ \kmsmpc,
$\Omega_{0}=0.25$, $\Omega_{\Lambda}=0.75$ is used throughout this
paper.  
We assume a Salpeter \citep{salp55} initial mass function (IMF) for all derived stellar masses and SFRs  and correct existing measurements used in this paper to a Salpeter IMF. 
We note that metallicity and abundance are taken to denote oxygen abundance, O/H, throughout this paper, unless otherwise specified.
In addition, we use dex throughout to denote the antilogarithm,
that is, 0.3\,dex is a factor of two.


\section{Data and measurements}
\label{sec:data}

\begin{figure*}
\includegraphics[width=7cm,angle=270,clip=true]{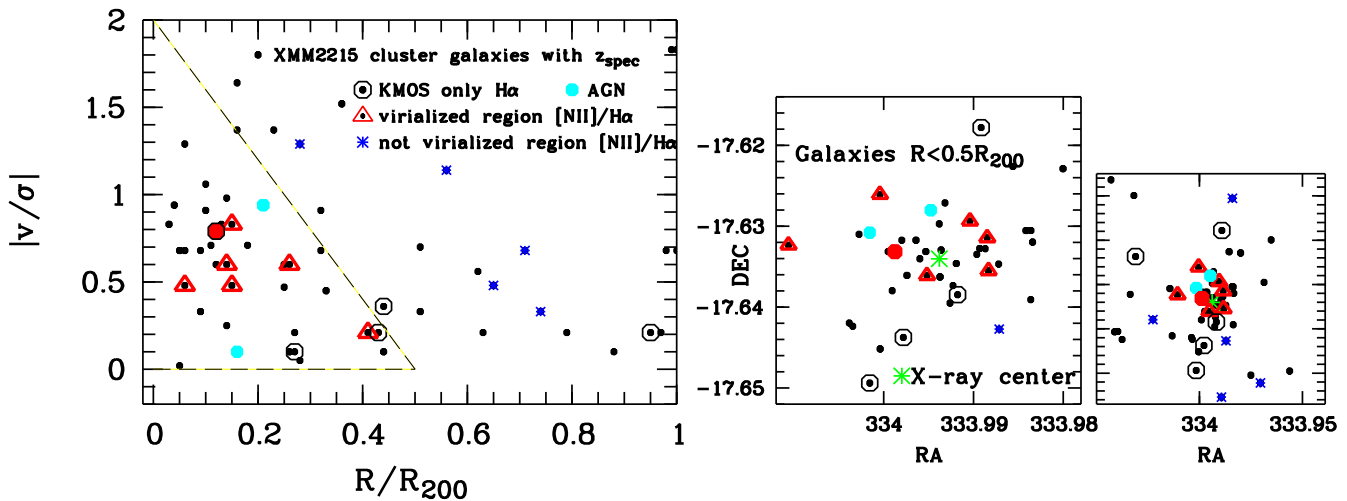}
\caption
{
\label{fig:PhaseSpace} 
\footnotesize 
Left: Phase-space diagram showing 58 XMM2215 cluster galaxies with spectroscopic redshifts inside $R_{200}$.  The galaxies with KMOS H-band observations are shown by larger symbols, as indicated in the legend. The dashed large triangle in the lower left-hand corner shows the virialized region, as derived by \citet{rhee17} using cosmological hydrodynamic simulations of groups and clusters. 
Center: Spatial distribution of 42 galaxies in the central part of the XMM2215 cluster at $R<0.5R_{200}$  (symbols like in the left panel). The red filled circle indicates one galaxy with a relatively high [NII]/\Ha\, ratio (but not as high as for the two objects shown in cyan) indicating possible AGN activity.
A green star symbol shows the cluster center determined with extended X-ray emission \citep{stanford06}.
Right: Spatial distribution of all 58 cluster galaxies shown in the phase-space diagram (left panel).
}
\end{figure*}


\subsection{KMOS observations and data reduction of cluster SF galaxies in XMM2215}

~~~The cluster XMM2215 is a massive $z \sim 1.46$ cluster discovered in the XMM Cluster Survey \citep{stanford06}. A previous estimation of the mass gave $M_{200} = 2.1^{-0.8}_{+1.9} \cdot 10^{14}M_{\odot}$ \citep{hilton10,stott10}, and a more recent estimation in this study gives  $M_{200} = (6.3 \pm 1.2) \cdot 10^{14}M_{\odot}$ (see Sect.\,\ref{sec:phasespace}). 
One advantage of observational environmental studies in this cluster compared to other clusters at $z>1$ is its wealth of spectroscopic redshifts published not only for SF but also for passive galaxies \citep{hilton09,hilton10,hayashi11,beifiori17,chan18}. This enables  the accretion state of cluster member galaxies to be characterized by identifying virialized and infalling regions (see Sect.\,\ref{sec:phasespace}).

The targets for the  KMOS H-band observations were selected from the list of $1.44<z<1.48$ [OII] emitters, with a flux larger than $2 \times 10^{-17} \rm{ergs/s/cm}^{2}$, identified from narrow-bands NB912 and/or NB921 by \citet{hayashi14}. Most of the targets had already been spectroscopically confirmed before the KMOS observations. 
VLT/KMOS H-band observations were carried out in 2016 on the night between October 19 and 20 (ESO program ID 098.A-0204(B)) with a nod-to sky strategy and each observing block (OB) consisting of  several ABA ABA sequences, where “A” signifies that the IFUs were placed on the science targets, and “B” indicates that the IFUs were observing blank sky. The integration time of each exposure was 300\,s with a total exposure time of 9900\,s. Seeing conditions measured in the optical by the DIMM seeing monitor ranged between 0.6 and 1.0 arcsec.

The KMOS data reduction was carried out with the official ESO-KMOS pipeline version 1.4.3 with the default settings of the pipeline, with the exception of the default of the pipeline to first combine exposures from a single OB before then combining these OBs into final data cubes. Since the best data quality in the co-addition of individual exposures is achieved using sigma clipping, we modified the default approach and ran the sigma clipping and cube combination on all 33 individual exposures (from four OBs) at once.


\subsection{KMOS emission-line measurements}

~~~To measure the main ELs, \Ha\, and \NII,\, from the KMOS H-band observations, we use
the 
Interactive Data Language (IDL) publicly available software KUBEVIZ \citep{fossati16} applied to each extracted 1D spectrum. 
To measure the ratio of the \Ha\, and the (typically) fainter \NII\, EL flux to compute O/H metallicities, we extracted 1D spectra summed over 25 spaxels centered on the  \Ha\, EL of the respective KMOS data cube, corresponding to an aperture of about 1 square arcsecond.
This typically corresponds to the area where KUBEVIZ measures  $S/N>3$ per spaxel in \NII.
For the total \Ha\, fluxes used to compute SFRs we extracted 1D spectra over slightly larger areas, including all spaxels where KUBEVIZ measures  $S/N>3$ in \Ha.
Due to differences in resolution in each IFU \citep{davies13}, the instrumental resolution at the wavelength of \Ha\, was computed from the skylines for every IFU in order to estimate the line widths of \Ha\, and \NII.
Starting from an initial redshift, KUBEVIZ  simultaneuosly fits the ``lineset'' \Ha\, and \NII. This ``lineset'' is described by a
combination of 1D Gaussian functions keeping the velocity separation of the lines fixed according to the line wavelengths (see Fig.\,\ref{fig:HaNII}). The continuum level is evaluated inside two symmetric windows between 80 and 200\AA\, redward and blueward of each line, omitting regions with other emission lines and using only values between the 40th and 60th percentiles. 
During the fit, KUBEVIZ takes into account the noise from the noise data cube, thus optimally suppressing sky line residuals.
We additionally fitted the \Ha\, and \NII\, lines including also the second (weaker) \NIIbl\, line. We checked that the fit of three ELs produces \Ha\, and \NII\, flux values in agreement (given the error bars) with the \Ha\, and \NII\, values using fits of two lines. Because the fainter \NIIbl\, line is more often heavily affected by night-sky lines than the brighter \NII, we decided to use the measurements of \Ha\, and \NII\, fluxes from the fit of two lines.

From the 20 observed XMM2215 targets we could detect the \Ha\, EL for 19 galaxies (see Table \ref{MeasXMM2215}).
The \NII\, EL could be detected in twelve galaxies, while in the remaining seven galaxies [NII] was either affected by a strong OH line or was too faint to be detected, yielding only an upper limit in two cases. 
For one galaxy among the seven, the \Ha\, line was detected but was heavily affected by a strong night-sky line, and therefore no reliable measurement of the \Ha\, EL flux was possible.
Two of the twelve galaxies with detected [NII] turn out to be AGNs because they have an \NII\, flux comparable with the \Ha\, flux (see Table \ref{MeasXMM2215}).
The remaining ten galaxies with detected [NII] include seven galaxies in the inner region of the cluster ($R<0.5R_{200}$) and three galaxies outside it. We note that two additional galaxies with upper limits on [NII] (see Table\,\ref{MeasXMM2215}) are also shown in Fig.\,\ref{fig:PhaseSpace} with blue symbols. 

\subsection{Stellar masses, SFRs, and oxygen abundances}
\label{SFRsox}

~~~ Stellar masses of the XMM2215  cluster galaxies were calculated using the code 
\emph{Lephare} of \citet{arnilb11}, which fits 
stellar population synthesis
models \citep{bruzcharl03} to the available SUBARU B, $R_{c}$, i', z', and WFCAM/UKIRT K-band photometry described in \citet{hayashi14}.
This code is a simple
$\chi^{2}$ minimization algorithm that finds 
the best match of templates for the given data. 
We do not fit templates with stellar ages of less than 0.5\,Gyr and larger than 5\,Gyr. This is a valid assumption since 
at $z \sim 1.5$ the Universe is only $\sim$4.5 Gyr old.
In addition, we  kept the redshift fixed, limited the number of extinction $E_{(B-V)}$ values (0 to 0.5, in 0.1 steps), and 
used a \citet{calzetti00} extinction curve.
We are confident in the robustness of the calculated masses since 
the $z-K$ color encompasses the redshifted 4000\AA\, break and thus is sensitive to galaxy mass-to-light ratios \citep{kaufm03}.

It turns out that one galaxy (39683, red filled circle in Fig.\,\ref{fig:PhaseSpace}) of our sample of ten SF galaxies with detected [NII]
has a high log([NII]/\Ha)$=-0.20 \pm 0.06$ ratio indicating possible AGN activity.
  For our study of environmental effects on metallicities we therefore restrict our cluster sample  to  nine galaxies with [NII]/\Ha\, measured and $\rm{log}(M/M_{\odot})<11$, including six objects at $R<0.5R_{200}$ (our \emph{SFvirialized} sample), but we also show the O/H measurement for 39683 in the MZR diagram, although with a different symbol (red filled circle).

~~~Since we do not have measurements of \Hb\, we cannot use an observed Balmer decrement to derive extinction. Therefore, the \Ha\, line luminosities $\rm{L}(\rm{H}\alpha)$ were corrected for three values of assumed extinction $A_{V}=0,0.5,1$  and then transformed into SFRs by applying  the \citet{ken98} conversion: $\rm{SFR} (M_{\odot}\rm{yr}^{-1}) = 7.9 \times 10^{-42} \rm{L}(\rm{H}\alpha)\rm{ergs/s}$.

~~~To estimate the chemical abundances, a number of 
diagnostics have been developed based on strong ELs, 
 ${\rm [O\,II]\,}{\lambda\,3727}$, H$\beta$, ${\rm [O\,III]\,}{\lambda\,5007}$, H$\alpha,$ and [NII]${\lambda\,6584}$.
At higher redshifts, these ELs move to the NIR, and studies of metallicities up to $z \sim 2.5$ use mostly the N2 calibration (or sometimes the O3N2 calibration) of \citet{petpag04} to derive oxygen abundances. For easier comparison with existing publications of the MZR at $z>1$ we therefore use the N2 
metallicity calibration for this study.  
We derive oxygen abundances from [NII]/\Ha\,using the N2 method of \citet{petpag04}. We note that one galaxy from our \emph{SFvirialized} sample has an [OIII]/\Hb\, measurement (with large error bars) and another galaxy has an upper limit for [OIII]/\Hb, both from the work of \citet{hayashi11},  and both indicating a ratio of [OIII]/\Hb\, smaller than one.

\section{Results}

\subsection{Spatial distribution and phase-space diagram}
\label{sec:phasespace}


\begin{figure}
\includegraphics[width=6cm,angle=270,clip=true]{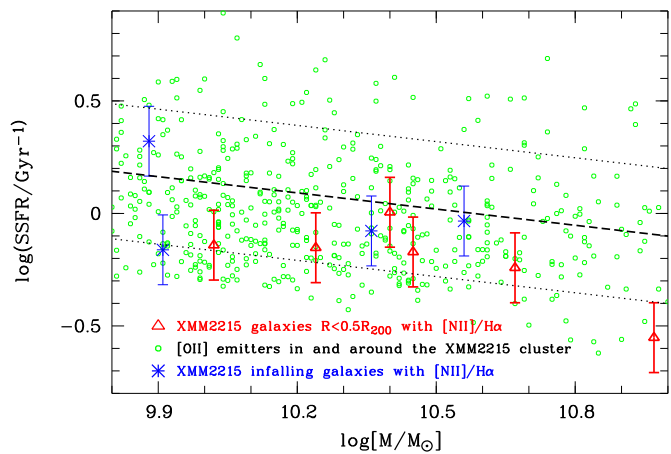}
\caption
{
\label{fig:SSFRXMM2215} 
\footnotesize 
Mass$-$SSFR relation for the [OII] emitters in and around the XMM2215 cluster (green open circles), the six XMM2215 cluster galaxies of the \emph{SFvirialized} sample (red triangles) and four infalling cluster galaxies (blue symbols).   The oblique solid line shows the   MS at $z \sim 1.46$ and its dispersion (indicated by the dotted lines) using Eq.\,1 in \citet{peng10}.
}
\end{figure}


\begin{figure*}
\includegraphics[width=7cm,angle=270,clip=true]{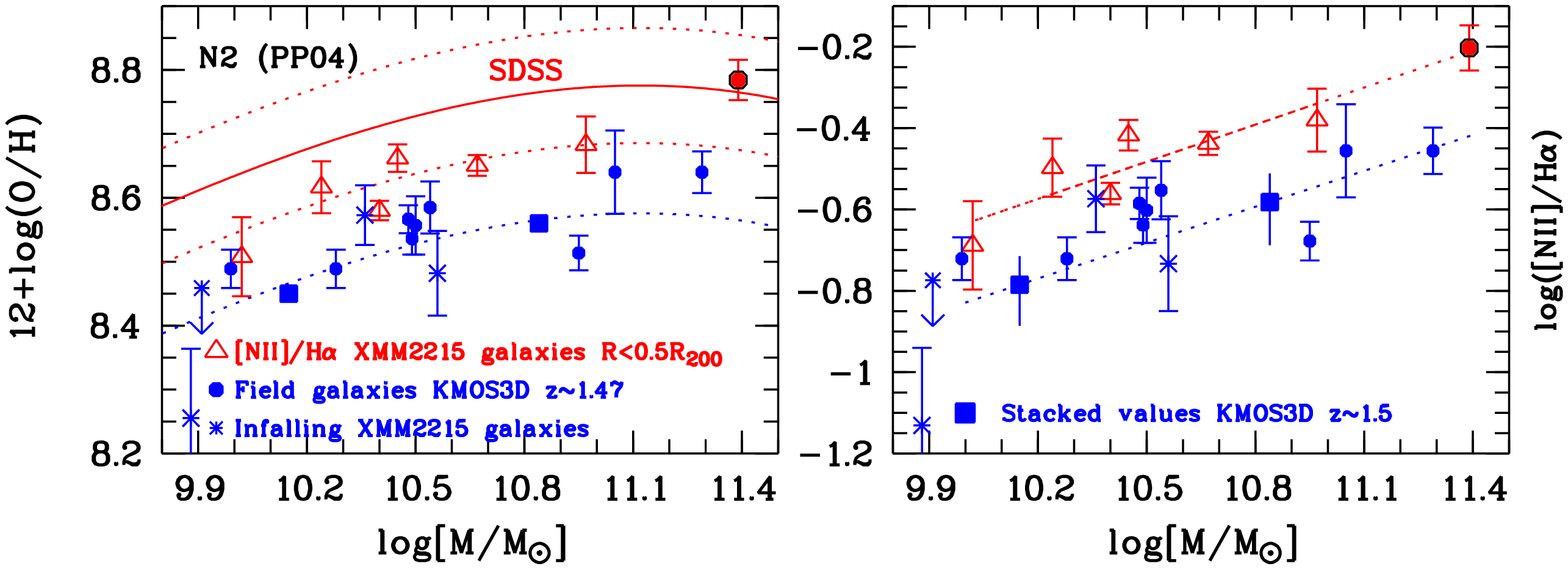}
\caption
{
\label{fig:MZR2215} 
\footnotesize 
Left: MZR for cluster and field galaxies at $z\sim 1.5$, using the N2 method of \citet{petpag04} to derive oxygen abundances.
The SDSS local MZR relation is shown by the red solid line using the fit given by  \citet{keweli08} for the N2 calibration. The dotted red lines show the dispersion of the SDSS relation, $\pm 0.09$dex, as given by \citet{keweli08}, while the dotted blue line is displaced by a further 0.11\,dex to lower metallicities compared to the lower dotted red line, and corresponds to the location of the stacked measurements (large blue filled squares) of a larger sample of $z \sim 1.5$ KMOS3D field galaxies from \citet{wuyts16}. 
Red triangles are XMM2215 cluster galaxies from the \emph{SFvirialized} sample, blue star symbols are XMM2215 infalling galaxies (see phase-space diagram, Fig.\,\ref{fig:PhaseSpace}), and blue filled circles depict KMOS3D field galaxies at $1.42<z<1.52$ from \citet{wuyts16}. 
The red filled circle is the galaxy 39683 with a relatively high [NII]/\Ha\, ratio, and a possible contribution from an AGN to the [NII] emission.
An enhancement of  the oxygen abundances of about 0.1\,dex for XMM2215 cluster galaxies compared to field is seen.
Right: Line ratios \NII/\Ha\, for the \emph{SFvirialized} sample (red), infalling galaxies, and KMOS3D field galaxies (blue). A linear fit of the log(\NII/\Ha) vs. stellar mass relation for the \emph{SFvirialized} sample is shown by the red dotted line, and  the linear relation implied by the stacked values of the KMOS3D sample (blue large filled squares) is shown by the blue dotted line. An enhancement of  the log(\NII/\Ha) ratios  of 0.2\,dex for XMM2215 cluster galaxies compared to field is visible.
}
\end{figure*}

~~~A phase-space diagram is a useful tool to characterize the accretion state of cluster member galaxies 
relatively free from effects due to the 2D projected
positions with respect to the cluster center (see Fig.\,\ref{fig:PhaseSpace}, comparing the three panels).
To investigate cluster membership for XMM2215 we collected redshift information from the literature \citep{hilton09,hilton10,hayashi11,beifiori17,chan18} yielding a sample of 108
galaxies with published spectroscopic redshifts. We then used the $3 \sigma$-clipping technique of \citet{yahvid77}, which assumes that clusters are relaxed isothermal spheres and that the velocity distribution of cluster galaxies follows an underlying Gaussian distribution.  Out of the 108 galaxies, 74 lie inside $3 \sigma$.

A mass model of the cluster was computed assuming that the cluster is a singular isothermal sphere.
$R_{200}$ 
is roughly equivalent to the virial radius of \citet{carlberg97}
and is calculated as $\sqrt(3)/10\cdot \sigma_{z}/H(z)$, where $\sigma_{z}$ is the velocity dispersion of the cluster and H(z) the redshift dependent Hubble parameter.
From this iterative method based on the technique of \citet{carlberg97} we obtained $R_{200} = (1.23 \pm 0.18)Mpc$ and $M_{200} = (6.3 \pm 1.2) \cdot 10^{14}M_{\odot}$.
The uncertainties of $M_{200}$ and $R_{200}$ were determined by randomly taking out ten galaxies of the spectroscopic sample of 108 galaxies, and then repeating the phase-space analysis a few hundred times.
Our derived $M_{200}$ and $R_{200}$  are slightly higher than those derived by \citet{hilton10} using a similar analysis ($M_{200} = 2.1^{-0.8}_{+1.9} \cdot 10^{14}M_{\odot}$, $R_{200} = (0.8 \pm 0.1)Mpc$), but it should be noted that \citet{hilton10} used only 44 spectroscopic redshifts (compared to 74 used here), and
we also used spectroscopic redshifts of passive galaxies more recently published by \citet{beifiori17} and \cite{chan18}.

From the 74 objects with spectroscopic redshifts inside $3 \sigma$ we found 58 galaxies to lie inside $R_{200}$, as depicted in Fig.\,\ref{fig:PhaseSpace}.
The red triangle symbols represent our main \emph{SFvirialized} sample of cluster galaxies for which [NII]/\Ha\, has been measured and which lie in the virialized region of the cluster, as indicated by the large triangle in the lower left-hand corner derived by \citet{rhee17} using cosmological hydrodynamic simulations of groups
and clusters.


\subsection{Star-forming XMM2215 cluster galaxies with enhanced metallicities}

~~~Figure\,\ref{fig:SSFRXMM2215} shows the mass--SSFR relation of SF
galaxies in and around the XMM2215 cluster. \citet{peng10} derived a formula of the evolution of the SSFR as a function of mass and time that we use to calculate the mean SSFR as a function of stellar mass at $z \sim 1.5$ (main sequence, MS). For this, we assume a dependence of the SSFR on mass as observed for local SDSS galaxies, an increase in the mean SSFR from $z \sim 0 $ to $z \sim 1.5$ as derived by \citet{peng10}, and a dispersion (indicated by the dotted oblique lines) of a factor of 0.3\,dex about the mean relation.
Star-formation rates for [OII] emitters, identified from narrow-bands NB912 and/or NB921 by \citet{hayashi14}, are derived from their [OII] fluxes using the relation recommended by \citet{maier15}. 
For the six galaxies in the $SFvirialized$ sample  and the four infalling cluster galaxies we assumed three values of extinction, $A_{V}=0,0.5$, and 1,  and then transformed their \Ha\, fluxes into SFRs by applying  the \citet{ken98} conversion. The error bars of the six red triangles and of the blue symbols in Fig.\,\ref{fig:SSFRXMM2215} indicate the range of SFRs for $A_{V}$ values between 0 and 1.
We use the \citet{calzetti00} $E_{B-V} = A_{V}/3.1$ (or 4.05) formula and their 
$E_{B-Vstellar} / E_{B-Vemission} = 0.44$ conversion between the color excess of the stellar continuum $E_{B-Vstellar}$ and the color excess derived from the gas emission lines $E_{B-Vemission}$.  Using  the minimum derived $E_{B-Vstellar}=0.2$ value for our \emph{SFvirialized} galaxies we obtain  $A_{V}=1.4$ (or $A_{V}=1.8$). We note that our assumed $A_V=0$, 0.5, and 1 values are not higher than  or in contradiction with the $E_{B-Vstellar}$ values.

~~~Five out of the six $SFvirialized$ galaxies lie in the lower SSFR part of the MS region, while the most massive one lies just below the MS region assuming a maximum of $A_{V}=1$; this latter would however lie in the lower SSFR part of the MS region for a plausible higher extinction value at higher masses of $A_{V}>1$. Therefore, we conclude that the six $SFvirialized$ galaxies are still  forming stars, although with slightly lower SSFRs, lying in the lower SSFR part of the MS region. We note that their slightly lower SFRs and higher metallicities (as discussed below) are in relatively good agreement with expectations  of the link between SFRs and O/Hs described by the fundamental metallicity relation \citep[e.g.,][]{mannu10}. For the four galaxies outside the virialized region (blue symbols) no clear trend of high/low SSFR is seen, due to the small number statistics.


~~~The left panel of Fig.\,\ref{fig:MZR2215} shows the MZR, using the N2-calibration of \citet{petpag04}, of the six $SFvirialized$  cluster galaxies (triangles) compared to the infalling galaxies (blue asterisks), and the local SDSS MZR (red solid line). To compare our sample with a field sample at similar redshifts with similar data quality, we used field galaxies at 
$1.42<z<1.52$ with KMOS H-band spectroscopy from the KMOS3D survey,  with [NII]/\Ha\, measurements published by \citet{wuyts16} shown as filled blue circles. We note, from looking at Fig.\,2 of \citet{wuyts16}, that their $z \sim 1.5$ stacked [NII]/\Ha\,  measurements seem consistent with $z \sim 1.5$ literature results from the FMOS-COSMOS survey \citep{zahid14}.
The dotted red lines show the dispersion of the SDSS relation, $\pm 0.09$dex, as given by \citet{keweli08}, while the dotted blue line is displaced by further 0.11\,dex to lower metallicities compared to the lower dotted red line, and corresponds to the location of the stacked measurements of a larger sample of $z \sim 1.5$ KMOS3D field galaxies (large blue filled squares).
Because the MZR of cluster galaxies inside $0.5R_{200}$ is roughly consistent with the SDSS position of the lower dotted red line,
the offset between the blue dotted line and the lower red dotted line indicates an increase in metallicities by a factor of 0.11\,dex for \emph{SFvirialized} cluster galaxies compared to field galaxies.

~~~Due to concerns surrounding the N2-metallicity indicator (see Sect.\,\ref{sec:intro}), we also show our results in the right panel of Fig.\,\ref{fig:MZR2215} in terms of the observed \NII/\Ha\, ratios.
We performed a linear fit of the log(\NII/\Ha)-versus-stellar mass relation for the \emph{SFvirialized} sample (red dotted line), and show the linear relation implied by the stacked values of the KMOS3D sample as a blue dotted line.
The log([NII]/\Ha) offset between the blue and red dotted lines is 0.2\,dex.
Given the mean uncertainties of the stacked values from \citet{wuyts16} of about 0.07\,dex, we conclude that this enhancement in log([NII]/\Ha) ratios has a 2.9$\sigma$ significance.
Thus, Fig.\,\ref{fig:MZR2215} clearly reveals an enhancement of the log(\NII/\Ha) ratios by 0.2\,dex and of the derived oxygen abundances by  0.11\,dex for XMM2215 cluster galaxies inside $0.5R_{200}$ compared to field. The different values for the enhancement (0.2\,dex vs. 0.11\,dex) reflect the factor 0.57 in the N2 conversion, $12+log(O/H)=8.90+0.57 \cdot log([NII]/\Ha)$.  
%
%

\section{Discussion and Summary}
\label{sec:disc}

\subsection{Slow quenching implied by enhanced metallicities of cluster galaxies at $z \sim 1.5$}

~~~The enhanced metallicities of $z \sim 1.5$ cluster galaxies inside $0.5R_{200}$ compared to $z \sim 1.5$  field galaxies can be interpreted as a quenching indicator based on the comparison of observed quantities with metallicity-SFR-mass bathtub model predictions with inflowing gas \citep{lilly13} performed for lower-redshift clusters by \citet{maier16,maier19}. 
The interpretation is that the gas metallicities increase because their ISM is no  longer diluted by the inflow of pristine gas after the hot halo gas reservoir has been stripped in the cluster environment, likely around $0.5R_{200}$, initiating strangulation.

The enhancement of O/Hs of cluster galaxies compared to field galaxies is
in agreement with the enhancement in metallicities found by \citet{shimakawa15} in a protocluster at $z>2$ compared to field galaxies based on stacked measurements of [NII]/\Ha\, for protocluster galaxies.
 \citet{shimakawa15} also discussed  one strangulation scenario based on their result: for infalling cluster galaxies, 
once an infalling galaxy is incorporated into a common cluster halo, 
the gas reservoir of the galaxy may be stripped or truncated due to the interactions with the cluster environment.
This can not only expel low-metallicity gas trapped in the outer region of the galaxy, but also terminate the fresh pristine gas accretion on to the galaxy, increasing the retained gas metallicity.
On the other hand, the studies of \citet{kacprzak15} and \citet{tran15} for cluster galaxies at $z>1.6$ used the N2 calibration from stacking and found a similar MZR for cluster and field galaxies. Since the clusters studied by \citet{kacprzak15} and \citet{tran15} have a much lower total mass than our studied XMM2215 cluster, it is possible that environmental effects are not strong enough in these two clusters to produce an observable effect on metallicities, due to a possible overly low ICM density, as discussed below.

~~~The current KMOS study of cluster galaxies in XMM2215 shows, using {individual} metallicity measurements, that environmental effects on metallicities implying quenching are already starting to act at $z>1$.
\citet{maier19} discussed how the hot gas reservoir and the cold gas in the disk may be influenced for infalling galaxies into clusters, depending on the strength of the cluster potential and the ICM density.
Gas can be removed from infalling galaxies if the ram pressure exceeds the restoring force per unit area (gravitational restoring pressure) exerted by the galaxy, as first derived by \citet{gunngott72}.
XMM2215 contains only a few known passive galaxies for which spectroscopic redshifts have been determined \citep{beifiori17,chan18}.
Therefore, it is plausible that the ICM density in this cluster implies a ram pressure higher than the restoring pressure of the {hot} halo gas reservoir of cluster galaxies, but not higher than the restoring pressure of the {cold} gas in the disk out of which stars are currently born.

~~~Compared to the slow-then-rapid quenching scenario discussed by \citet{maier19} and \citet{roberts19}, we think that we see only the ``slow'' quenching in XMM2215 SF cluster galaxies at $z\sim 1.5$ traced by enhanced metallicities. As discussed in \citet{maier19} based on comparisons with high-resolution cosmological hydrodynamic simulations of \citet{bahe13}, the ram pressure is probably only slightly larger than $3 \times 10^{-14} N/m^{2}$ in XMM2215,
larger than the restoring pressure of hot gas and enough to strip the more extended, less dense, and less tightly bound hot gas. 
This slow quenching is also in relatively good agreement with results of \citet{hayashi17,hayashi18} who used ALMA to measure CO(2-1) 
ELs and dust continuum and derived molecular gas masses for several XMM2215 cluster galaxies. These latter authors discussed that the main component of the stripped gas is the neutral gas reservoir, 
while the molecular gas (observed with ALMA) is relatively much less affected by the ram pressure. This means that the galaxies continue to form stars from the molecular gas, in agreement with our slow quenching scenario.
The ICM density in XMM2215 has probably not yet reached the threshold found by \citet{roberts19} to initiate the ``rapid'' quenching, a ram pressure threshold required to strip the cold gas of cluster galaxies; cold gas which is denser and sits much closer to the galactic center.


\subsection{Summary}

~~~To study environmental effects on the chemical enrichment and quenching of cluster galaxies, we used  KMOS spectroscopy of \Ha\, and \NII\, in the massive cluster XMM2215 at $z \sim 1.5$ to measure oxygen abundances and estimate SFRs. 
Using the phase-space diagram results of Fig.\,\ref{fig:PhaseSpace} we identified XMM2215 SF cluster galaxies in a virialized region with $R<0.5R_{200}$ and smaller line-of-sight velocities. 
Studying the metallicities of a sub-sample of the cluster galaxies with KMOS H-band observations and located in the virialized region  (\emph{SFvirialized} sample), we found evidence for enhanced metallicities by about $0.1$\,dex for these galaxies compared to infalling and field galaxies at similar redshifts (Fig.\,\ref{fig:MZR2215}).
These cluster galaxies are still forming stars, although at slightly lower SFRs (see Fig.\,\ref{fig:SSFRXMM2215}).

These findings indicate a strangulation scenario in which the ICM density toward the cluster center becomes high enough such that RPS can remove the hot halo gas reservoir of cluster galaxies, thus enhancing metallicities and initiating slow quenching, while the galaxies continue to form stars using cold gas in the disk and enhance their metallicities because no fresh pristine gas accretion dilutes their ISM.
Additional spectroscopic observations, not only of [NII] and \Ha\, but also of [OIII] and \Hb,\, for a  larger sample of cluster galaxies at $z>1$ are needed to reinforce these interesting 
results, which for now are based on a small sample.

\begin{acknowledgements}
We would like to thank the anonymous referee for providing constructive comments and help in improving the manuscript. 
\end{acknowledgements}


\end{document}